\documentclass[aps,prb,twocolumn,10pt,superscriptaddress]{revtex4-2}

\usepackage{header}
\begin{document}
\title{
Near-deterministic photon entanglement from a spin qudit in silicon using third quantisation
}
\author{G\"ozde \"Ust\"un}
\affiliation{School of Electrical Engineering and Telecommunications, UNSW Sydney, Sydney,
NSW 2052, Australia}
\affiliation{ARC Centre of Excellence for Quantum Computation and Communication Technology, Melbourne, VIC, Australia}
\author{Samuel Elman}
\affiliation{Centre for Quantum Software and Information, University of Technology Sydney, Sydney, New South Wales 2007, Australia}
\author{Jarryd Pla}
\affiliation{School of Electrical Engineering and Telecommunications, UNSW Sydney, Sydney,
NSW 2052, Australia}
\author{Andrew Doherty}
\affiliation{School of Physics, The University of Sydney, Sydney, NSW 2052, Australia}
\author{Andrea Morello}
\affiliation{School of Electrical Engineering and Telecommunications, UNSW Sydney, Sydney,
NSW 2052, Australia}
\affiliation{ARC Centre of Excellence for Quantum Computation and Communication Technology, Melbourne, VIC, Australia}
\author{Simon J. Devitt}
\affiliation{Centre for Quantum Software and Information, University of Technology Sydney, Sydney, New South Wales 2007, Australia}
\affiliation{InstituteQ, Aalto University, 02150 Espoo, Finland}
\floatsetup[figure]{style=plain,subcapbesideposition=top}

\begin{abstract}

Unlike other quantum hardware, photonic quantum architectures can produce millions of qubits from a single device. However, controlling photonic qubits remains challenging, even at small scales, due to their weak interactions, making non-deterministic gates in linear optics unavoidable. Nevertheless, a single photon can readily spread over multiple modes and create entanglement within the multiple modes deterministically. Rudolph’s concept of third quantization leverages this feature by evolving multiple single-photons into multiple modes, distributing them uniformly and randomly to different parties, and creating multipartite entanglement without interactions between photons or non-deterministic gates. This method requires only classical communication and deterministic entanglement within multi-mode single-photon states and enables universal quantum computing. The multipartite entanglement generated within the third quantization framework is nearly deterministic, where “deterministic” is achieved in the asymptotic limit of a large system size. In this work, we propose a near-term experiment using antimony donor in a silicon chip to realize third quantization. Utilizing the eight energy levels of antimony, one can generate two eight-mode single-photon states independently and distribute them to parties. This enables a random multipartite Bell-state experiment, achieving a Bell state with an upper-bound efficiency of 87.5\% among 56 random pairs without non-deterministic entangling gates. This approach opens alternative pathways for silicon-based photonic quantum computing.
\end{abstract}
\maketitle
\section{Introduction}
In the pursuit of fault-tolerant quantum computers, two families of architectures have emerged: matter-based~\cite{Leek_2007, Chatterjee_2021, stuyck2024cmoscompatibilitysemiconductorspin} and photonic-based hardware~\cite{alexander2024manufacturable, Maring2024}. These two types have distinct advantages and challenges, leading to separate pathways for fault-tolerant architectures. In matter-based architectures (such as superconducting or spins in semiconductors), the approach is to build
arrays of physical qubits with precise control~\cite{Stemp2024, M_dzik_2022, Krinner_2022}. Controlling small numbers of matter-based qubits has been achieved with great success because the qubits in these architectures interact with each other and the gates are, in principle, deterministic~\cite{Krinner_2022, Tanttu_2024, M_dzik_2022, holmes2023improved}. At present, the challenge in building a utility-scale machine lies in scaling the number of qubits into the millions.
Photonic architectures present the opposite challenge: while a single device can produce millions of photons, the real difficulty lies in controlling them, even at small scales~\cite{alexander2024manufacturable}. Photons have extremely weak interactions with one another, with the most common gates being non-deterministic, measurement-induced coupling. Even in an ideal, error-free scenario, the success of these gates cannot achieve unit efficiency. As such controlling photons, particularly
making them interact and creating probabilistic entanglement, is very difficult to scale.

However, one remarkable property of photons is their ability to create a certain type of entanglement deterministically; such as a single photon spread over many modes. Such entanglement may enable quantum computation; however, purely linear optical approaches that use only deterministic entanglement of photons for universal quantum computation encounter severe scaling challenges, as either the circuit complexity increases exponentially with the number of photons~\cite{Reck94}, or they fail to provide a universal platform for quantum computing~\cite{aaronson2010computationalcomplexitylinearoptics}.
Consequently, there has been no scalable universal quantum computing approach relying solely on deterministic entanglement. To address this, Rudolph~\cite{rudolph2021everywhere} proposed producing many single photons, each splitting into multiple modes. These modes are then distributed randomly and uniformly to different parties, creating multipartite entanglement. This idea introduces a new concept called third quantization~\cite{third_quantization_note}, in which non-interacting photons can encode first quantization through second quantization, creating entanglement without interactions between individual photons or non-deterministic gates. Under third quantization, universal quantum computing emerges purely from the deterministic entanglement of multi-mode single-photon states and classical communication. -- where the latter is used after each party receives their modes. Notably, in this case, the number of elements required and the local dimensionality grow polynomially. Unlike other paradigms such as Fusion-Based Quantum Computation~\cite{bartolucci2021fusionbasedquantumcomputation} or Measurement-Based Quantum Computation~\cite{jozsa2005introductionmeasurementbasedquantum,Briegel2009}, the third quantization framework does not require multi-photon entangled states (or, more formally, graph states) as input; instead, it relies solely on independent single photons and outputs highly entangled states.

\begin{figure*}[htb!]
        \includegraphics[width=.8\linewidth]{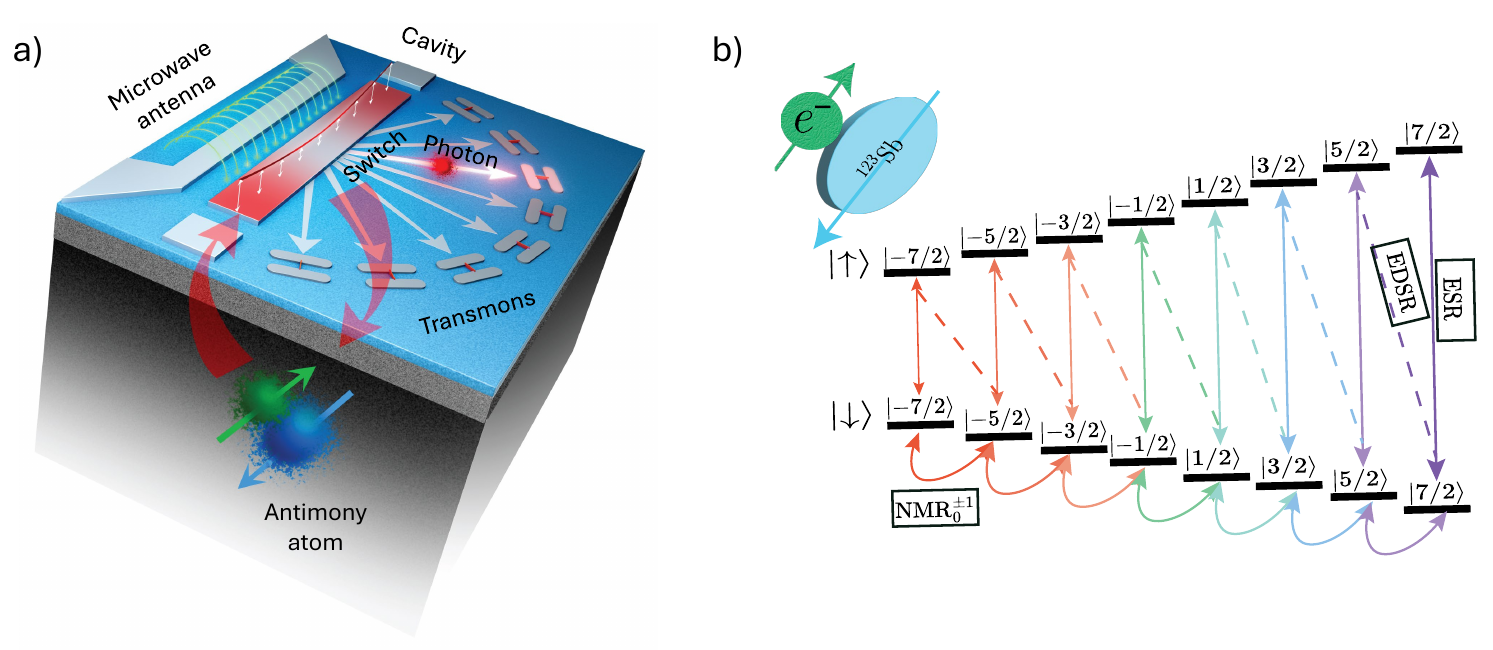}
    \captionsetup{font=small}
    \caption[Proposed Device for Third Quantisation Experiment and Antimony Donor]{ \label{fig:antimonydonor} \textbf{Proposed Device for Third Quantisation Experiment and Antimony Donor} a): Artist's impression of the random, multipartite Bell state device. The device has a microwave antenna to drive EDSR, ESR and NMR transitions. Additionally, a cavity, placed near the microwave antenna, is coupled with the electric dipole of the donor electron as shown with red arrows. The antimony donor is assumed positioned at an antinode of the cavity electric field, to achieve capacitive coupling to the donor's electric dipole and stimulate the EDSR transitions to emit photons coherently. (see details in~\ref{total_H} for the total Hamiltonian of the system.) A microwave switch, represented by gray arrows, sends the emitted photon, the red circle, to one of the transmon qubits for readout. b) The energy spectrum of the neutral antimony donor. Antimony is a high-spin nuclei and it has 8 energy levels. When the antimony donor is in its neutral charge state, the spin of the electron couples with the spin of the antimony nucleus, resulting in a doubled Hilbert space dimension, which is a 16-level system. Curved arrows represent $\rm{NMR}_0^{\pm 1}$ transitions, while ESR is depicted using vertical solid arrows, and EDSR is indicated with dashed arrows. The $0$ subscript in NMR emphasise that the antimony donor is in its neutral charge state. 
    } 
\end{figure*}

In this work, we propose a realistic experiment for the simplest realization of third quantization by using a single high-spin antimony donor in silicon. 
Using antimony to independently generate two multi-mode single photon states and distribute them to parties, we propose a random multipartite Bell state experiment that relies solely on third quantization, requiring no non-deterministic gates. 
This proposed experiment can produce a Bell state among 56 random pairs of recipients with an upper bound efficiency of 87.5\%. Bell states are the simplest stabiliser states and useful for measurement-based quantum computing.  To go beyond Bell experiment and perform universal quantum computation with near-deterministic efficiency, this approach can be extended to larger systems with multiple donors, which are high-dimensional. 
Our approach goes beyond simply proposing the third quantization framework: it can be extended to generate more complex states than third quantization requires, thereby enabling new possibilities in measurement-based donor architectures—such as constructing high-dimensional cluster states.
Our proposed experiment is also suitable for other types of semiconductor qubits. With this work, we hope to reconsider alternative paths that are suitable for photons, to scalable quantum computing.
\section{Antimony donor}\label{sec_antimony} Our proposed silicon quantum device employs antimony (${^{123}}$Sb) donor with a high nuclear spin, $I=7/2$. Therefore, the nuclear spin alone creates a $2I+1=8$-dimensional Hilbert space.  As a group V element - having five valence electrons - ${^{123}}$Sb is a hydrogen-like impurity which, at low temperatures, binds an extra electron(Fig.~\ref{fig:antimonydonor}b). In this situation, we refer to it as a neutral donor, i.e. the donor is in the neutral charge state. It is in general possible to remove the extra electron, thus leaving the donor in the charge-positive, ionized state, but for the purpose of this work we wish to have the electron bound to the donor in order to provide the microwave-frequency spin transitions that inject photons in the microwave cavity. 
Because of the extra electron, the resulting system spans a 16-dimensional Hilbert space \cite{fernandez2024navigating}. 

Through ion implantation, a single antimony atom can be introduced into the silicon lattice, displacing one of the native silicon atoms. The antimony isotope $^{123}$Sb exhibits a nuclear spin (I) of $7/2$ and a gyromagnetic ratio ($\gamma_n$) of 5.55 MHz/T. The nucleus's non-spherical charge distribution results in an electric quadrupole moment $q_n = [-0.49, -0.69] \times 10^{-28} \, \text{m}^2$. The donor-bound electron, possessing a spin (S) of $1/2$, has a gyromagnetic ratio ($\gamma_e$) approximately equal to 27.97 GHz/T. It is magnetically coupled to the nuclear spin through the Fermi contact hyperfine interaction $A \hat{S} \cdot \hat{I}$, where A is 101.52 MHz in bulk silicon. 

Compared to the other high spin donor candidates in group-V such as $^{209}$Bi and $^{75}$As, ${^{123}}$Sb has a several advantages. While bismuth has a higher nuclear spin, $I=9/2$, its large mass results in excessive lattice damage, and low activation probability. Additionally, bismuth has a much higher hyperfine coupling ($A=1475.4$~= MHz) compared to antimony ($A=101.52$~MHz), which introduces a strong  measurement back-action on the nuclear spin during the readout process \cite{joecker2024error}. Arsenic, in contrast, has a lower nuclear spin, $I=3/2$, which makes it unsuitable for utilizing the third quantization formalism to its full potential.

The spin Hamiltonian of the antimony atom is described in a 16-dimensional Hilbert space, shown in Fig~\ref{fig:antimonydonor}b, built from the tensor product of the electron spin operator and two nuclear spin operators \cite{fernandez2024navigating}: 
\begin{equation}\label{sb}
\begin{aligned}
    H_{\text{Sb}} = B_0(-\gamma_n \hat{I_z} + \gamma_e \hat{S_z}) + A(\Vec{S}\cdot\Vec{I}) + \sum_{\alpha, \beta \in \{x,y,z\}} Q_{\alpha \beta}\hat{I}_\alpha \hat{I}_\beta
\end{aligned}    
\end{equation}
The first term in this Hamiltonian accounts for the Zeeman splitting, on both the electron and the nuclei. The second term describes the hyperfine interaction that arises from the overlap of electron and nuclear wavefunctions. In the last term, where $\alpha, \beta = {x, y, z}$ represent Cartesian axes, $\hat{I}_\alpha$ and $\hat{I}_\beta$ are the corresponding 8-dimensional nuclear spin projection operators. The term $Q_{\alpha\beta} = e q_n V_{\alpha\beta}/2I(2I-1)h$ represents the nuclear quadrupole interaction energy, determined by the electric field gradient (EFG) tensor $V_{\alpha\beta} = \partial^2V (x, y, z)/\partial\alpha\partial\beta$ \cite{asaad2020coherent}. This quadrupole interaction introduces an orientation-dependent energy shift to the nuclear Zeeman levels, enabling the individual addressability of nuclear states. 
In Equation~\eqref{sb}, $B_0$ represents the magnetic field in which the nanoelectronic device containing the antimony donor is placed, with a value approximately equal to $1 \text{T}$.
This ensures that the eigenstates of $H_{\text{Sb}}$ are approximately the tensor products of the nuclear states $\ket{m_I}$ with the eigenstates $ \{\ket{\downarrow} , \ket{\uparrow} \} $ of $\hat{S}_z$ because $\gamma_e B_0 \gg A \gg Q_{\alpha\beta}$. The latter condition implies $H_{\text{Sb}} \approx B_0(-\gamma_n \hat{I_z} + \gamma_e \hat{S_z}) + A(\Vec{S} \cdot \Vec{I})$ ensuring that the nuclear spin operator approximately commutes with the electron-nuclear interaction.
This condition allows for an approximate quantum non-demolition (QND) readout of the nuclear spin via the electron spin ancilla \cite{joecker2024error}.

Coherent transitions between the $^{123}$Sb spin eigenstates can be induced by magnetic and electric fields, on both the electron and the nuclear spin~\cite{fernandez2024navigating}. Electron spin resonance (ESR), which flips the spin of the electron, is achieved by the driving term $H^{\text{ESR}} = B_1 \gamma_e \hat{S}_x \cos(2\pi f_{m_I}^{\text{ESR}}t) $. Here, $B_1$ is the amplitude of an oscillating magnetic field at one of the eight resonance frequencies $f_{m_I}^{\text{ESR}}$ determined by the nuclear spin projection $m_I$. Nuclear magnetic resonance (NMR), which changes the nuclear spin projection by one quantum of angular momentum, is achieved by the driving term $H^{\text{NMR}} = B_1 \gamma_n \hat{I}_x \mathrm{cos}(2\pi f_{m_I - 1 \leftrightarrow m_I}^{\text{NMR}}t)$. In addition to magnetically driven transitions (ESR, NMR), there exists the possibility of inducing spin transitions electrically. This is accomplished by leveraging the combined states of the electron and nucleus, representing a high-spin extension of the `flip-flop' transition recently demonstrated in the $I = 1/2$ $^{31}$P system~\cite{Rosty}. The application of an oscillating electric field $E_1 \cos\left(2\pi f_{m_I-1\leftrightarrow m_I}^{\text{EDSR}} t\right)$ leads to electric dipole spin resonance transitions (EDSR) in the neutral donor. This process dynamically modulates the hyperfine interaction $A(E_1) \hat{S}_{\pm} \hat{I}_{\pm}$ through the Stark effect~\cite{tosi_2017}, where the $\pm$ subscripts denote the rising and lowering operators, respectively. This mechanism conserves the total angular momentum of the combined electron-nuclear states. Consequently, the EDSR transitions manifest as diagonal (dashed) lines in Fig.~\ref{fig:antimonydonor}b.

Our proposed device features a silicon nanoelectronic component fabricated on top, as depicted in Fig~\ref{fig:antimonydonor}a. This component includes a broadband microwave antenna that delivers the $B_1$ field required for ESR and NMR. To drive the donor spins electrically at microwave frequencies via EDSR, we can either leverage the stray electric fields generated by the microwave antenna \cite{fernandez2024navigating}, or fabricate a dedicated open-circuit antenna \cite{Rosty}.

\section{Experimental protocols for controlling multi-modes photonic states}\label{sec_experiment}

\begin{figure*}
    \includegraphics[width=.6\linewidth]{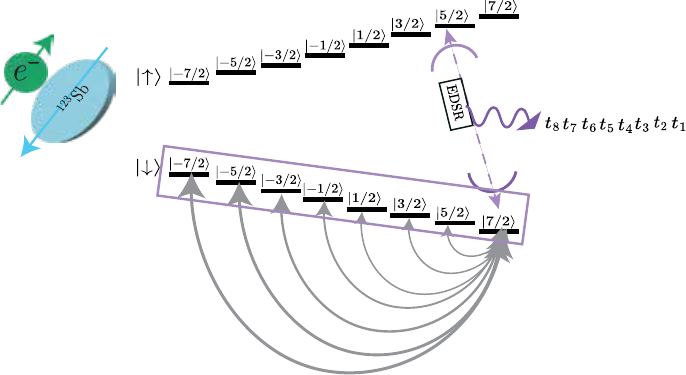}
    \caption[Time-bin protocol for controlling multi-mode photonic states]{\label{fig:multimodcontrol}\textbf{Time-bin protocol for controlling multi-mode photonic states} The EDSR frequency between the states $\ket{7/2}\ket{\downarrow} \leftrightarrow \ket{5/2}\ket{\uparrow}$ is chosen as fixed frequency for emitting photons. The purple rectangle represents the uniform superposition. The gray arrows represent the `permutation operation' operation between nuclear spin states. $t_1$, $t_2$, $\cdots$, $t_8$ represent the time-bins into which the photon is emitted. (see details in~\ref{time-binsteps}.)}
\end{figure*}
Our goal is to encode a single photon into multiple modes and  thus create high-dimensional entanglement deterministically. To achieve this, we propose the integration of microwave cavities into the device, from which a photon is emitted, employing either frequency multiplexing or time-bin multiplexing. In frequency multiplexing, a single photon is emitted in a superposition of different frequencies belonging to different microwave cavities. This requires either: a) using as many microwave cavities as the frequencies the system has; or, b) a tunable cavity that can be tuned for the different range of frequencies. This method is not feasible for near-term experiments (see details in~\ref{freq-mult}.) Instead, we can use time-bin multiplexing to generate deterministic entanglement in time at a fixed frequency and employ the cavity only in this frequency( see the total hamiltonian of the system in~\ref{total_H}).  With time-bin multiplexing, engineering concerns related to cavity utilization, tuning speed for a broad frequency range, and similar considerations are obviated.

\subsection{Time-bin multiplexing} 
In this scenario, the time-bin into which the photon (of a single fixed frequency) is emitted represents the designated parameter, see Figure~\ref{fig:multimodcontrol}. 
In the case of EDSR, the coupling occurs through an electric dipole, which enables a faster emission process compared to ESR, where the coupling occurs through a magnetic dipole. This is because the electric dipole \cite{tosi_2017,Rosty} provides a much stronger coupling to the cavity field than the magnetic dipole~\cite{Tossi_2014}. 
In this protocol, starting from the state $\ket{\psi_0} = \ket{7/2}\ket{\downarrow}\ket{vac}$, (where $\ket{vac}$ represents the cavity vacuum state), we prepare the uniform superposition of the nuclear spin states (for example, by applying a qudit hadamard operation):
\[
    \begin{split}
        \ket{\psi_1}=\frac{1}{\sqrt{8}}\big(& \ket{7/2} + \ket{5 /2} + \ket{3/2} \\&+ \ket{1/2} + \ket{-1/2} + \ket{-3/2} \\&+ \ket{-5/2} + \ket{-7/2} \big) \ket{\downarrow}\ket{vac},
    \end{split}
\]
where the electron is in the spin-down state $\ket{\downarrow}$. An EDSR pulse is then applied, which flips the spin of the electron conditioned on the nucleus being in the state $\ket{7/2}$. The electron experiences a coherent exchange of energy with the microwave cavity such that at a time $t_1$ the electron is in the down state and the cavity is populated with a photon of frequency ${\omega_1}$. 
A permutation operation, to swap the population between nuclear states, is then applied between $\ket{7/2}$ and $\ket{5/2}$ with the electron in $\ket{\downarrow}$, leaving the system in the state:
\[
    \begin{split}
        \ket{\psi_2}=&\frac{1}{\sqrt{8}}(\ket{5/2} \ket{\downarrow}\ket{\omega_1}_{t_1}+(\ket{7/2}+\ket{3/2}\\&+\ket{1/2}+\ket{-1/2}+\ket{-3/2}\\&+\ket{-5/2}+\ket{-7/2})\ket{\downarrow}\ket{vac})
    \end{split}
\]
This process --- an EDSR pulse, the electron/cavity interaction, photon emission, and permutation operation --- is repeated. As a result, a single photon is spread over different time-bins $t_i$. The final state is then the coherent superposition:
\begin{equation} \label{eq:sbphoton}
    \begin{aligned}
       \ket{\psi_{\text{final}}} =\frac{1}{\sqrt{8}}(\ket{5/2}\ket{\downarrow}\ket{\omega_1}_{t1} + \ket{3/2}\ket{\downarrow}\ket{\omega_1}_{t2}\\+ \ket{1/2}\ket{\downarrow}\ket{\omega_1}_{t3}+ \ket{-1/2}\ket{\downarrow}\ket{\omega_1}_{t4}\\+\ket{-3/2}\ket{\downarrow}\ket{\omega_1}_{t5}+\ket{-5/2}\ket{\downarrow}\ket{\omega_1}_{t6}\\
        +\ket{-7/2}\ket{\downarrow}\ket{\omega_1}_{t7} + \ket{7/2}\ket{\downarrow}\ket{\omega_1}_{t8})
    \end{aligned}
\end{equation}
The resultant, $\ket{\psi_{\text{final}}}$, is still not a pure photonic state. It exhibits entanglement with the antimony donor. To decouple the photon from the antimony, a qudit Hadamard operation is applied to the states of antimony in Eq~\ref{eq:sbphoton}. Subsequently, the nuclear states are measured (see details in~\ref{decouplingSb}). This configuration results in a $W$ state, a high-dimensional entangled state created deterministically by encoding the photon into time bins. The resultant state becomes $\ket{\psi_{\text{final'}}} = 
\ket{W_8}$  where  $\ket{W_8}  = \ket{\omega_1}_{t_1} + \ket{\omega_1}_{t_2} + \ket{\omega_1}_{t_3} + \ket{\omega_1}_{t_4} + \ket{\omega_1}_{t_5} + \ket{\omega_1}_{t_6} + \ket{\omega_1}_{t_7} + \ket{\omega_1}_{t_8}$ is a $W$ state that exhibits entanglement arising from the temporal times at which photons are emitted. The time-bins are the encoding parameter for photon emission and can be represented in binary form. Specifically, $ \ket{vac}$  is $\ket{0}^{\otimes 8}$ represents the absence of a photon and corresponds to a vacuum state. On the other hand, $\ket{\omega_1}_{t_1} = \ket{10000000}$ represents the emission of a photon at time $t_1$, $ \ket{\omega_1}_{t_2} = \ket{01000000}$ represents the emission of a photon at time $t_2$, and so forth. 

The application of the permutation operation between $\ket{7/2}$ and $\ket{5/2}$ is trivial and corresponds to a single NMR $\pi$-pulse, which typically takes only a few tens of microseconds~\cite{fernandez2024navigating, Yu2025}. The operation can be applied in sequence to other NMR resonances to swap the populations of the $\ket{7/2}$ state with states other than $\ket{5/2}$. Another method for the application of `permutation operation' is to perform a sub-global rotation~\cite{Yu2025}: The process involves applying the ESR transition to the states we want to swap, bring them to the spin-up state of the electron, and swapping them via a global rotation~\cite{Yu2025}. It enables the swapping of any states of antimony without affecting the population of other states. Then, we can bring the swapped states back to the $\ket{\downarrow}$ state of the electron and proceed with photon emission through the cavity. The ESR pulse typically lasts about one microsecond, while swapping the targeted states via global rotation can take up to a few hundreds of microseconds. The choice between the two methods depends significantly on their respective speeds. Notably, for every pulse applied, increasing the $B_1$ magnetic field in the driving term results in even shorter durations \cite{Pla_2012}. 

Readout: One physical realization that will be possible in the near future, involves using a microwave photon counter at millikelvin temperatures, with a superconducting resonator (qubit) for readout, as described by \citet{Wang_2023}.  
Subsequently, partial states corresponding to each temporal iteration can be transmitted. For example, at time $t_1$, the state $\ket{5/2}\ket{\downarrow}\ket{w1}_{t_1}$ can be dispatched for readout. In the absence of the photon, a partial collapse of the wave function ensues; conversely, its presence prompts a complete collapse. 

While this operational paradigm pertains to the near future, the envisaged progression involves the deployment of eight superconducting qubits instead of one. These qubits are meticulously adjusted to span distinct time steps, facilitating the storage of photons in their respective superconducting qubits until the eighth superconducting qubit is reached at the end of the temporal sequence.

Error localization is categorized into errors from the Jaynes-Cummings component and the antimony Hamiltonian component, where actions like permutation operations are conducted. Photon emission is a coherent process governed by the Jaynes-Cummings model, and the time ($t_e$) required to emit a photon is dependent upon the coupling strength. In the case of EDSR, the coupling strength between the cavity and the electric dipole is found to be 3 MHz in~\cite{tosi_2017}, so that the photon emission time is approximately $1/3$ MHz, or around 333~ns. 
The $T_1$ relaxation time for the electron spin in antimony is several seconds ($2.44\,\text{s}$~\cite{fernandez2024navigating}). \color{black} Consequently, when emitting a photon, the system will not transition to any state other than the one associated with the cavity’s specific configuration. Other dominant error sources in the context of spin coupling with a cavity is spin dephasing, which occurs at the rate $1/T_2^*$ for the corresponding transition with which the cavity is in resonance. The $T_2^*$ for the electron is found to be $510 \mu$s for the $m_I = \ket{7/2}$~\cite{fernandez2024navigating}. Therefore, the photon generation rate (3 MHz)  is three orders of magnitude faster than the dephasing rate of the antimony electron (1.96 kHz). \color{black}Photon losses resulting from the finite quality factor (Q) of the cavity, occurring at a rate of $\omega/Q$ where $\omega$ is the frequency of the cavity~\cite{wyatt_2,Wyatt_1,Samkharadze_2018,Osika_2022}. This is discussed in the next section. Apart from the spin coupled with the cavity, there is another aspect where we can encounter potential error sources: the application of operators. This includes applying ESR pulses, EDSR pulses, and permutation operations. Within the spin system, the predominant error source is the phase flip.
The operations and their average success are shown in the following table:

\begin{table}[htb]
\centering
\resizebox{\columnwidth}{!}{
\begin{tabular}{|c|c|}
\hline
\textbf{Operations within the antimony donor} & \textbf{Fidelity, \% \cite{Yu2025, fernandez2024navigating}} \\
\hline
State Initialization &  99.5\\
NMR pulses (or single qubit nuclear gate)  &  99.8 \\
ESR  pulses &   99.5 \\
EDSR pulses &  99.5 \\
\hline
\end{tabular}
}
\caption{Average success rate of operations within antimony, as demonstrated experimentally~\cite{asaad2020coherent,fernandez2024navigating,Yu2025}}
\label{tab:antimony_success}
\end{table}
The simplest permutation gate is between \(\ket{7/2}\) and \(\ket{5/2}\), corresponding to a single NMR pulse with a fidelity of 99.8\%. 

As shown in the table, there are, of course, some associated error rates; however, antimony is an extremely well-characterized physical system. Its Hamiltonian is now known to a precision of a few hertz~\cite{asaad2020coherent}, and its noise is extremely low, with both the noise characteristics and the environment being well understood~\cite{asaad2020coherent,fernandez2024navigating,Yu2025}. An additional significant advantage of using antimony donors as quantum emitters lies in the nature of the donor-bound electron, which is weakly coupled to the nucleus. This feature enables the simultaneous emission of photons once multiple emitters are integrated. For instance, when two antimony donors are implanted into a silicon lattice, each possesses its own bound electron, allowing the simultaneous emission of two photons. More generally, implanting \( n \) antimony atoms enables the simultaneous emission of \( n \) photons—provided that \( n \) individual cavities are fabricated to couple to each emitter.  In systems with multiple donors, slight variations in the EDSR transition frequencies can arise, primarily due to fabrication imperfections. Factors that can lead to differences include: 
\begin{itemize}
    \item Hyperfine value: This depends on the donor’s position relative to the gates and the exact gate structure, which can vary slightly due to fabrication imperfections. Strain differences can cause small changes in the hyperfine value, typically within a few MHz, but larger variations may occur if the donor is close to the interface.
    \item Quadrupole term: Also strain-dependent, this varies with donor location and gate structure. Observed values range from ~4 kHz to ~50 kHz. 
\end{itemize}

However, the same strong-coupling regime (typically a few MHz) can be maintained even with multiple donors, because the coupling strength is determined by the derivative of the hyperfine interaction with respect to the electric field rather than by its absolute value. Small variations in the photon emission frequency can still arise, since the cavity is tuned to the EDSR transition, leading to very small frequency differences between emitted photons. This issue is expected to be mitigated as fabrication technologies continue to improve. For instance, while timed-ion implantation does not provide precise control over donor placement, deterministic ion implantation enables accurate control of the donor’s position and proximity to the surface. In addition, techniques have been developed to enhance placement precision even within timed-ion implantation schemes~\cite{holmes2023improved}. A more detailed discussion of these effects can be found in~\cite{ustun2025comparingschemescreatingqudit}.

\color{black}Decoherence: We can emit a photon in approximately 333 ns through EDSR, while the electron $T_2^*$ time is 510 $\mu$s. The emission of photons can persist until the electron experiences dephasing. ESR (EDSR) pulses, have a duration of approximately 1 $\mu$s ( 10 $\mu$s). It is imperative to acknowledge the presence of the nuclear component, utilized for creating a general superposition and `permutation' operations. Application of a qudit Hadamard may take up to 100 $\mu$s and application of permutation operations may take up to a few hundreads of $\mu$s. The $T_2^H$ time for the nucleus is found to be $247 \mu$s in~\cite{fernandez2024navigating}.
Nonetheless, notable advancements on the hardware front are underway to extend the $T_2$ times, including improved enrichment of the spin-free $^{28}$Si isotope \cite{acharya2024highly}, enhanced readout methodologies \cite{santiagosito}, and the design of novel microwave antennas \cite{dehollain2012nanoscale} to enhance system control efficiency. Furthermore, employing time-bin multiplexing protocol allows us to create a general superposition not only within the electron $\ket{\downarrow}$ manifold but also within the electron $\ket{\uparrow}$ manifold. In other words, the system can accommodate more than eight levels, leading to a higher-dimensional $W$ state, contingent upon the coherence time of antimony.

\section{Near-deterministic photon entanglement using antimony donor via third quantisation}\label{sec_3rdQuantization} In this section, we explain the third quantization formalism and how the \textit{W} state, created by the time-bin protocol, will be utilized within this formalism.

The third quantization method, is a mathematical framework introduced by Rudolph~\cite{rudolph2021everywhere}, \color{black} that creates entanglement without requiring the direct interaction between individual photons. The method uses completely independent multi-mode single photon states. By distributing  multi-mode single photon states among many parties, we create multipartite entanglement along with a fully symmetric subspace over the modes.
In the third quantization framework, one can define a specific type of measurements in which each party performs independent measurements on the subset of modes they hold, followed by feedforward (i.e., outcome-dependent interferometers on the remaining unmeasured modes prior to their measurement)~\cite{rudolph2021everywhere} (also mentioned in Appendix for completeness - \ref{background_3rdq}). For example, if each party holds four photons in total, they may choose to measure only the first photon while leaving the remaining three untouched. All photons undergo identical measurement procedures at each step. In most systems, this process yields a rank $>$ 1 outcome, leaving the measured photons still entangled with the unmeasured ones~\cite{rudolph2021everywhere}.\color{black} Using only local operations and classical communication, third quantization enables a variety of protocols, including universal quantum computing.~\cite{rudolph2021everywhere, Knill2001}. 

As we state in the Supplementary Information and explicitly shown by Rudolph in~\cite{rudolph2021everywhere}, a third-quantized state can be constructed by preparing \( N \) copies of the \( \ket{W_K} \) state, where \( N \) is the number of photons and \( K \) is the number of modes (or parties). Each mode in the \( \ket{W_K} \) state is then distributed to a separate party. In this construction, the expansion of \( \ket{W_K}^{\otimes N} \) approximates a third-quantized state, with the approximation error scaling as \( \mathcal{O}(1/N) \) as \( N \rightarrow \infty \). This means the error can be made arbitrarily small by increasing the number of photons. To ensure the error remains sufficiently small, at least \( K = N^3 \) parties are required.

\color{black}The objective of this proposal is to realize the simplest version of a third-quantized state $\ket{\Sigma}$. In its most basic form, this requires at least two photons ($N=2$), ensuring that the dimension of $W$ state is at least 8, as determined by $K = N^3$.
As mentioned in the previous section, one can create the $\ket{W_8}$ state by using antimony and time-bin multiplexing. 
Therefore, the number of parties, denoted as $K$, is 8. 
Each mode in state $\ket{W_8}$ goes to a party as follows:
\begin{equation}
   \small
    \begin{aligned}
        \ket{W_8} = \frac{1}{\sqrt{8}}( &\mid\overset{\overset{\text{A}}{\downarrow}}{1}\underset{\underset{\text{B}}{\downarrow}}{0}\overset{\overset{\text{C}}{\downarrow}}{0}\underset{\underset{\text{D}}{\downarrow}}{0}
        \overset{\overset{\text{E}}{\downarrow}}{0}\underset{\underset{\text{F}}{\downarrow}}{0}\overset{\overset{\text{G}}{\downarrow}}{0}\underset{\underset{\text{H}}{\downarrow}}{0}\succ + \mid01000000\succ + \mid00100000\succ + \\ &\mid00010000\succ + \mid00001000\succ + 
        \mid00000100\succ +\\ &\mid00000010\succ + \mid00000001\succ
        )
    \end{aligned}
\end{equation} Here, we follow Rudolph's notation and use $|.\succ$ for second quantisation. In this equation, the letters represent the parties. For example, 1st mode goes to party A, second mode goes to party B and so on. This equation is equivalent to the following form:
\begin{equation}
   \small
    \begin{aligned}
        \ket{W_8} = \frac{1}{\sqrt{8}} ( \!
&\mid\!1\!\succ_A\mid\!0\!\succ_B\mid\!0\!\succ_C\mid\!0\!\succ_D\mid\!0\!\succ_E\mid\!0\!\succ_F\mid\!0\!\succ_G\mid\!0\!\succ_H + \\
&\cdots + \\
&\mid\!0\!\succ_A\mid\!0\!\succ_B\mid\!0\!\succ_C\mid\!0\!\succ_D\mid\!0\!\succ_E\mid\!0\!\succ_F\mid\!0\!\succ_G\mid\!1\!\succ_H )
    \end{aligned}
\end{equation} where each party has one mode. When two independent single photons are distributed among the parties, the resulting state will be $ \ket{W_8} \otimes \ket{W_8} $, or simply $\ket{W_8}^{\otimes 2}$, as follows:
\begin{equation}\label{eq:2W}
  \small
  \begin{aligned}
         \ket{W_8}^{\otimes 2} \approx \frac{1}{8}(
                        &\mid\overset{\overset{\text{A}}{\downarrow}}{1}\underset{\underset{\text{B}}{\downarrow}}{0}\overset{\overset{\text{C}}{\downarrow}}{0}\underset{\underset{\text{D}}{\downarrow}}{0}
        \overset{\overset{\text{E}}{\downarrow}}{0}\underset{\underset{\text{F}}{\downarrow}}{0}\overset{\overset{\text{G}}{\downarrow}}{0}\underset{\underset{\text{H}}{\downarrow}}{0}\succ \mid\overset{\overset{\text{A}}{\downarrow}}{0}\underset{\underset{\text{B}}{\downarrow}}{1}\overset{\overset{\text{C}}{\downarrow}}{0}\underset{\underset{\text{D}}{\downarrow}}{0}
        \overset{\overset{\text{E}}{\downarrow}}{0}\underset{\underset{\text{F}}{\downarrow}}{0}\overset{\overset{\text{G}}{\downarrow}}{0}\underset{\underset{\text{H}}{\downarrow}}{0}\succ + \\
                        &\mid10000000\succ\mid00100000\succ + \\
                        & \cdots + \\
                        &\mid00000001\succ\mid00000010\succ)
  \end{aligned}
\end{equation} Terms where the same party receives two photons are removed by post-selecting only outcomes in which photons are detected in distinct modes ($K=N^3$), leaving 56 terms. Since there are two photons in total, each party now receives two modes (one mode from each photon). Equation~\ref{eq:2W} can then be written as follows:

\begin{equation}\label{eq:shared_bell}
    \small
    \begin{aligned}
        \ket{W_8}^{\otimes 2} \approx 
       &\mid\!10\!\succ_A \bigg( \mid\!01\!\succ_B + \mid\!01\!\succ_C + \mid\!01\!\succ_D + \mid\!01\!\succ_E + \\
        &\mid\!01\!\succ_F + \mid\!01\!\succ_G + \mid\!01\!\succ_H \bigg) + \\
        &\mid\!10\!\succ_B \bigg( \mid\!01\!\succ_A + \mid\!01\!\succ_C + \mid\!01\!\succ_D + \mid\!01\!\succ_E + \\
        &\mid\!01\!\succ_F + \mid\!01\!\succ_G + \mid\!01\!\succ_H \bigg) + \\
        &\cdots + \\
        &\mid\!10\!\succ_H \bigg( \mid\!01\!\succ_A + \mid\!01\!\succ_B + \mid\!01\!\succ_C + \mid\!01\!\succ_D + \\
        &\mid\!01\!\succ_E + \mid\!01\!\succ_F + \mid\!01\!\succ_G \bigg)
    \end{aligned}
\end{equation} where parties holding vacuum are eliminated.
With appropriate grouping, the equation represents pairs of parties sharing a maximally entangled state, specifically a Bell state. This can be expressed as follows:
\begin{equation}
    \begin{aligned}
        \ket{W_8}^{\otimes 2} \approx \ket{\Psi_{AB}^{+}} + \ket{\Psi_{AC}^{+}}  + \cdots +  \ket{\Psi_{HF}^{+}} + \ket{\Psi_{HG}^{+}}
    \end{aligned}
\end{equation} where for $N=2$, the third quantized state $\ket{\Sigma}$ in Eq~\ref{thirdq} becomes $\ket{\Psi^{+}}$, a maximally entangled Bell state with a theoretical upper bound efficiency of 0.875 (see~\ref{sec:success}). 
In the experiment, each party performs its own measurement as if they had one qubit of a maximally entangled Bell state. Out of $K$ parties, $K - 2$ will detect vacuum. This protocol has a remarkable phase stability condition, as shown in~\cite{rudolph2021everywhere}. If each party applies a random phase rotation to their modes - for instance, party A has a random rotation $\theta_A$ and party B has $\theta_B$, and so on - this will not affect the efficiency of the protocol.

The schematic representation of the proposed device is shown in Figure~\ref{fig:antimonydonor}, where we use antimony donor coupled with a resonator to emit a photon at the EDSR frequency. After the photon is emitted from the cavity, each mode is directed to a party using a switch. The switch can be constructed from different materials, e.g., semiconductor switches~\cite{Pospieszalski} that operate over nanosecond timescales but which have sizable microwave losses (of approximately 2~dB), superconducting switches~\cite{Pechal_2016} that are compatible with current superconducting circuit fabrication techniques, or kinetic inductance switches~\cite{Wagner_2019} that operate with speeds in the sub-nanosecond range and with potentially low levels of loss~\cite{wyatt_2,Wyatt_1}.

Since uniary encoding is used in this protocol, energy is not conserved. For example, in the case of a basis choice involving the vacuum state, the vacuum is transformed into a superposition of the vacuum and $\ket{1}$ after the Hadamard operation, which does not conserve energy. This means that single photon unitaries directly on the photons are hard. To make this process physically feasible, each party is equipped with a transmon qubit. The operation is performed after the photon travels to and is absorbed by the transmon qubit, which is then used for readout. The proposed device has eight transmon qubits in total, representing the parties, and they are positioned on our silicon chip, which is located in a single dilution fridge.~(Figure\ref{fig:antimonydonor}).
We then allow a few tens of ns readout time for the superconducting qubit~\cite{Storz2023}. Regarding measurement errors at this step, the 
$T_1$ time for superconducting qubits can reach a few milliseconds~\cite{Bland2025}, which is much longer than the time required for photons to reach the transmons. Therefore, although the probability of a measurement error is not strictly zero, it is negligible compared to other error sources. 

However, it is important to note that our encoding differs from conventional single-rail encoding. In traditional single-rail encoding, photon loss directly results in logical errors, as the loss of a photon is indistinguishable from the logical \(\ket{0}\) state of a photon. In contrast, our protocol circumvents this issue. Since photons are emitted coherently and the entire process is clocked, photon loss can be reliably detected. A successful emission and detection event corresponds to a specific click pattern. If no click is registered at the expected time for any of the transmons, photon loss is inferred. Crucially, this loss does not propagate as a logical error within our protocol since we are able to detect the photon loss. This means that if no click is observed, the photon will be simply re-emitted.

Regarding photon emission: the two photons are emitted and measured sequentially, which is entirely appropriate, particularly for the \( N = 2 \) case. Within the third quantization framework, each party applies a unitary operation to their subset of modes. In this scenario, each party holds two modes, so a subset of two implies that each party measures one mode. Consequently, each measurement step destroys the photons, but the information is propagated to the next photon in the experiment. This process continues iteratively until the final output state is reached, at which point a measurement reveals the result of the experiment.


Regarding the photon loss estimation of the protocol: the use of a microwave cavity introduces two relevant quality factors:
\begin{itemize}
    \item $Q_i$: the internal quality factor of the cavity, which characterizes the rate at which photons are lost to the bath. Based on our previous findings \cite{vaartjes2023strong, Wyatt_1, wyatt_2}, this can be taken to be in the range $10^5$ to $10^6$.
    \item $Q_c$: the coupling quality factor, which characterizes the rate at which photons go to the detectors. This can be taken as $10^4$ \cite{vaartjes2023strong}.
\end{itemize}
The coupling rate of the cavity to the bath is: $\kappa_i = \frac{\omega_c}{Q_i}$ and the coupling rate of the cavity to the port which photons go to the detectors (in our case, the detectors are the transmons) is: $\kappa_c = \frac{\omega_c}{Q_c}$. The spin-photon coupling strength is, as noted in the previous section, found to be $g_s = 3\,\text{MHz}$ \cite{tosi_2017}. The overall rate at which photons go from antimony to the bath - resulting in photon loss - is: $\gamma_{\text{bath}} = \frac{g_s \kappa_i}{g_s + \kappa_i}$ and the overall rate at which photons go from antimony to the detectors is:$ \gamma_{\text{port}} = \frac{g_s \kappa_c}{g_s + \kappa_c}$

The ratio of photon loss is then: $ \text{Loss} = \frac{\gamma_{\text{bath}}}{\gamma_{\text{bath}} + \gamma_{\text{port}}}$ and the ratio of photons going to the detector, i.e. the fraction of photons reaching the detector, is: $\text{Success} = \frac{\gamma_{\text{port}}}{\gamma_{\text{bath}} + \gamma_{\text{port}}}$.
The success rate in dB is calculated as: $\text{dB} = 10 \log_{10}(\text{Success})$.
The EDSR frequency corresponding to the transition \( \ket{7/2}\ket{\downarrow} \leftrightarrow \ket{5/2}\ket{\uparrow} \) is 28.41~GHz~\cite{fernandez2024navigating}. Since the cavity is proposed to be placed on this EDSR transition for coherent photon emission, we may take \( \omega_c = 28.41\,\text{GHz} \). 
\begin{itemize}
    \item For $Q_i = 10^6$ (an upper bound), the overall loss is approximately $0.0189 \approx 2\%$, and the success rate is $0.981$, corresponding to $0.08\,\text{dB}$.
    \item In a very realistic, but a worse case scenario where $Q_i = 10^5$ ( a lower bound), the overall loss is $0.15$, and the success rate is $0.845$, corresponding to $0.7\,\text{dB}$.
\end{itemize} In previous work~\cite{vaartjes2023strong}, we achieved success rates higher than the lower bound considered above, with dB $=0.34$. However, to realistically simulate the effect of photon loss on the theoretical upper bound of the success probability, we consider a range of photon loss values from 1\% to 10\% per mode, capturing potential best- and worst-case scenarios. For the photon loss simulations: the system in the proposed experiment comprises a total of 16 modes. As we employ time-bin multiplexing scheme to emit photons from a single quantum emitter, without any delay lines, and each mode passes through the same physical component, we expect each mode to experience photon loss with approximately the same probability, $p$.

If the photon loss probability were exactly the same for both photons across all modes, its effect on the theoretical upper bound of the success probability would be given by $0.875(1 - p)^2$ where $p \in [0.01 , 0.1]$. Naturally, in an experimental setting, slight variations in loss between modes may still occur, even though photon loss is expected to be symmetric.

\begin{figure}[htb]
    \centering
    \includegraphics[width=\linewidth]{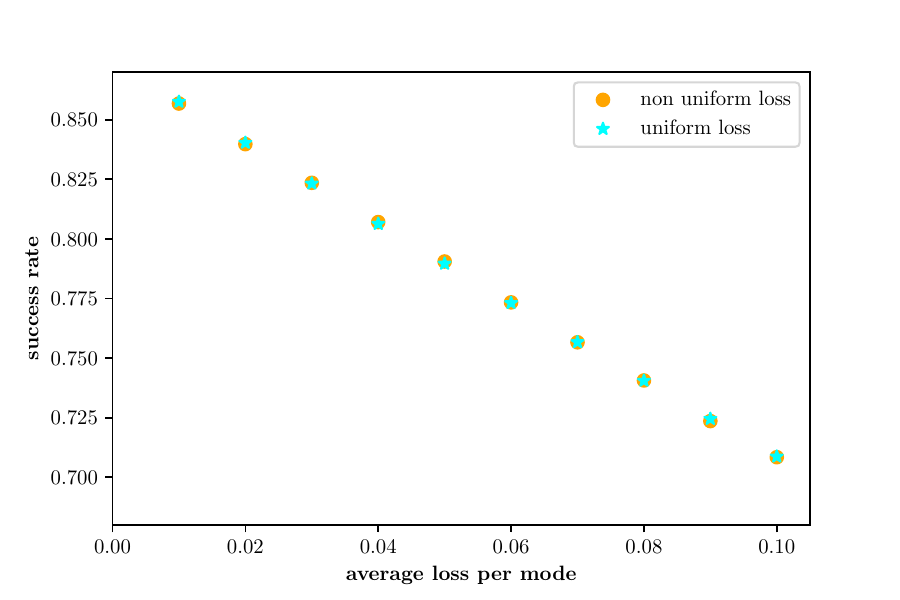}
    \caption[Effect of (non)uniform photon loss on success probability]{\textbf{Effect of (non)uniform photon loss on success probability} The total success rate is shown for two cases: when each mode has the exact same photon loss rate (blue stars), and when each mode has approximately the same loss rate, modeled as a normal distribution over the 16 modes with a standard deviation of 0.005 (orange dots).}
    \label{fig:non-uniform_photon_loss}
\end{figure}

\begin{figure} [h]
    \sidesubfloat[]{
    \includegraphics[width=.9\linewidth]{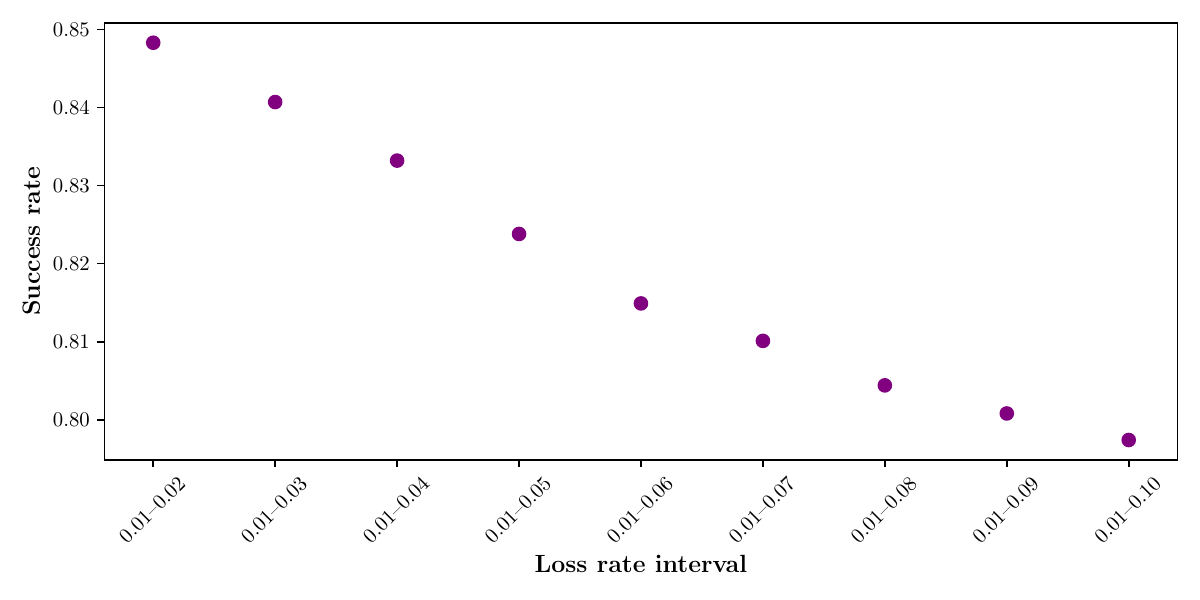}
    }
    
    \sidesubfloat[]{
    \includegraphics[width=.9\linewidth]{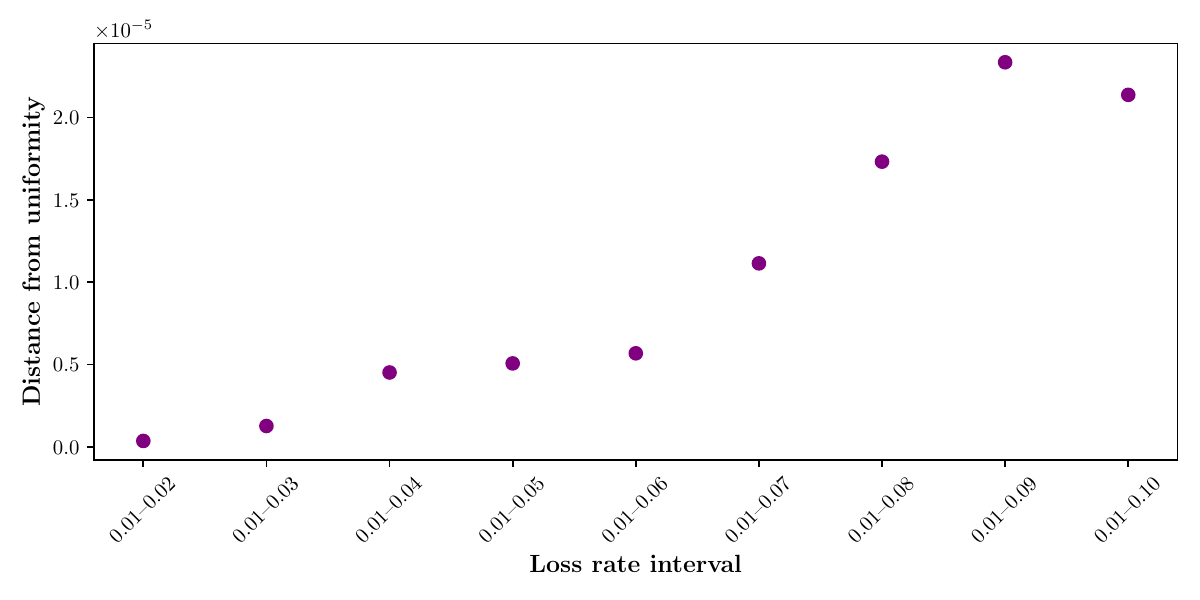}
    }
    \caption[Effect of random photon loss on success probability]{\textbf{Effect of random photon loss on success probability} a) The total success rate under random photon loss per mode. The x-axis indicates the interval within which each mode's loss rate is randomly selected. Compared to the previous plot, the photon loss is more randomized, but still follows a uniform distribution within the specified interval across the 16 modes.
    b)The distance from uniformity versus photon loss interval across modes is plotted. The distance from uniformity is calculated using the following formula:$\sum_i \left(q_i - \frac{1}{n_i}\right)^2$ where \(i\) indexes the successful patterns (i.e., cases in which no party receives more than one photon), \(n_i\) is the total number of successful patterns, and \(q_i\) is the normalized success probability corresponding to pattern \(i\).}
    \label{fig:random_photon_loss}
\end{figure}

Our simulations, which examine how photon loss impacts the theoretical upper bound of the protocol’s success probability, are presented in Figure~\ref{fig:non-uniform_photon_loss} and Figure~\ref{fig:random_photon_loss}.

Figure~\ref{fig:non-uniform_photon_loss} shows the case where the photon loss is approximately the same for each mode, with slight deviations. One can simulate the more irregular case where each mode can experience wildly different photon loss rate as shown by Figure~\ref{fig:random_photon_loss}. It is worth noting that photon loss in microwave resonators is generally lower than in optical systems~\cite{wallraff_2021,devoret_2013} and in general smaller than 10\%~\cite{vaartjes2023strong, magnard_2020}. 

\color{black}Using the third quantization formalism, we have proposed a random, multipartite Bell state experiment in which a Bell state is created between one of 56 random pairs with an efficiency of 0.875 by using a single high-dimensional antimony donor in a silicon chip.
Note that two single photon states ($\ket{W_8}$) are prepared completely independently and end up in different parties, ensuring that multi-photon interference plays no role in the protocol.

This proposed experiment is the simplest example of third quantization, where the third-quantized state (denoted as $\ket{\Sigma}$ in~\cite{rudolph2021everywhere} and in the Supplementary Information) becomes $\ket{\Psi^{+}}$, a maximally entangled Bell state.

Once scaled beyond two photons, the components of the experiment—such as the quantum emitters (antimony donors), transmon qubits for detecting microwave photons, and switches—remain the same; only their quantities increase. The minimal setup shown in Figure~\ref{fig:antimonydonor} includes all the essential components required for third quantization. 
If we wanted photons to reach the transmons simultaneously (which can be needed to go beyond random multipartite Bell game), one additional element required is delay lines for microwave photons. There has been recent progress in this direction. For instance, Ref.~\cite{Makihara2024ProgrammableDelay} reports the experimental realization of programmable delay lines for microwave photons using a Parametrically Addressed Delay Line (PADL). By driving parametric processes with pump fields, the signal’s propagation speed, direction, and coupling can be dynamically controlled. Although performance is not yet optimal, this approach represents an improvement over earlier implementations.

There is a fundamental difference between the third quantization protocol and approaches that generate graph states from quantum emitters, such as those in~\cite{pichler,Bartlett:21,Kianna}. At first glance, they may appear similar—indeed, the setup we propose is also capable of generating qudit graph states as suggested in~\cite{ustun2025comparingschemescreatingqudit}. However, the key distinction lies in how the photons are emitted. In third quantization, photons are emitted independently and do not interact with each other at any point in the protocol. In contrast, protocols for generating graph states emit photons in an entangled chain.

Although the \(N = 2\) case is a special scenario, the third quantization framework inherently relies on high-dimensional quantum states: each single photon is coherently spread across multiple modes, forming a high-dimensional state deterministically. The emitter itself is a high-dimensional quantum system, and for \(N > 2\), each party holds a high-dimensional subsystem of a single photon. By distributing the modes of a single photon to a party, we created a fully symmetric subspace over the modes which is not the case for other approaches~\cite{pichler,Kianna,Bartlett:21,Wallraff}.

The resultant states produced by third quantization for \(N > 2\) are fundamentally different from those generated in~\cite{Wallraff, Paeseni}, such as (high-dimensional) graph states or stabilizer states. Therefore, the framework of third quantization represents a completely different approach. 

\color{black}\section{Conclusion - Outlook}

In this work, we proposed an experiment where a single antimony atom, hosted in a silicon chip, is used to create a random multipartite Bell state via the framework of third quantisation. This experiment generates a Bell state between one of 56 possible pairs with an efficiency of 0.875 in the case of no experimental error. Furthermore, by increasing the number of parties, the success probability will improve even further. Notably, we created entanglement without using any non-deterministic gate, or photon-photon interactions. We started with non-interacting photons and ended up with maximally entangled Bell states. 

This proposal already outperforms standard methods, where two single photons are sent through two beam splitters and the success probability of obtaining a Bell state is 50\%. Beyond that, the proposal  has a single compact device, which, in principle can be scaled to more complex chipsets capable of universal quantum computing using the third quantisation model.  
Realization of third quantisation in a purely linear optical system poses challenges, specifically with the uniary encoding of the photonic states themselves - making necessary single qubit rotations difficult.

For future endeavors, perhaps this experiment could be done in a loophole-free manner. Then the critical timing must be considered, starting when the photons reach the transmon qubits for readout. If these transmon qubits are positioned in a single dilution fridge, the maximum distance between parties can be approximately 30 cm. To conduct this operation in a loophole-free manner, it is necessary to make the random basis choice and complete the readout within a few nanoseconds, ideally around 1 ns. Although the total time for choosing a random basis and reading out the state could be extended by increasing the distance between the parties, this remains infeasible for a single device within one dilution fridge. Using longer connections helps to increase the readout time, but in this situation, one would have to use more than one dilution fridge, as mentioned in ~\cite{Storz2023}.

To move beyond the Bell Game and fully demonstrate the potential of third quantization, it is necessary to increase the number of photons, and the dimensionality of the system that generates them. 
By using two antimony donor coupled with each other in a silicon chip, each with their own electrons, the Hilbert space dimension's will increase to 256 levels instead of 16. This is the natural extension of a setup already demonstrated with phosphorus atoms \cite{Stemp2024}. Consequently, the number of photons can be increased to six, enabling the implementation of a dual-rail Bell state~\cite{rudolph2021everywhere}, which requires four photons. Notably, for this experimental realization, the photons are not required to be identical~\cite{rudolph2021everywhere}. Using multiple antimony donors as quantum emitters also enables the simultaneous emission of multiple photons, as mentioned above. \color{black}Furthermore, we can start constructing the building blocks of fault-tolerant architectures for photonic hardware like in~\cite{bartolucci2021fusionbasedquantumcomputation, deGliniasty2024spinopticalquantum} for higher dimensions. For example, in the \( N = 3 \) case, the resulting state of third quantization is identical to the state onto which we project when performing qutrit fusion using a \(\ket{W}\) state ancilla~\cite{Luo,fusion_published}.

\color{black}Hence, it is worth noting that we primarily relied on a time-bin multiplexing scenario. However, once technology advances sufficiently to fabricate at least seven cavities on a single device, the frequency multiplexing protocol becomes highly advantageous, as it uses the system’s frequencies to encode a photon. This results in a polynomial decrease in the footprint in terms of time bins when such a scenario is employed.

\color{black}Another point to mention is that our goal here is not to claim this is extremely practical for photonic quantum computing architecture. Rather, it is to demonstrate that photons can generate useful high-dimensional entanglement by evolving into many modes.
This useful entanglement is powerful enough to perform universal computation in photonic architectures~\cite{rudolph2021everywhere}. Instead of following pathways designed for matter-based systems to create entanglement, we can explore different pathways for photons. What is useful and efficient for matter-based systems is not necessarily efficient and useful for photons. We hope that this work will open new avenues for understanding non-stabilizer codes and states, as well as their qudit versions in photonic hardware.

\section*{ACKNOWLEDGMENTs}
We thank Ryan Mann, Benjamin Wilhelm, Wyatt Wine and Jason Saied for their valuable discussions on third quantization, experimental limitations, microwave cavities and for reviewing the final section, respectively. Data regarding the loss calculation can be found in the auxiliary files on arXiv in version 2 of the manuscript~\cite{3rdQ2025Ancillary}.
This research was funded by the Australian Research Council Centres of Excellence for Quantum Computation and Communication Technology (CE170100012) and Engineered Quantum Systems. G.\"{U}. acknowledges support from the Sydney Quantum Academy. SJE was supported with funding from the Defense Advanced Research Projects Agency under the Quantum Benchmarking (QB) program under award no. HR00112230007, HR001121S0026, and HR001122C0074 contracts. The views, opinions and/or findings expressed are those of the authors and should not be interpreted as representing the official views or policies of the Department of Defense or the U.S. Government. J.J.P. acknowledges support from an Australian Research Council Future Fellowship (FT220100599). A.C.D. is supported by the Australian Research Council Centre of Excellence for Engineered Quantum Systems (EQUS, Grant No. CE170100009). A.M. acknowledges support from an Australian Research Council Laureate Fellowship (FL240100181). 

\onecolumngrid
\newpage
\section*{Appendix}


\setcounter{section}{0}
\setcounter{equation}{0}
\setcounter{figure}{0}

\renewcommand{\thesection}{A-\Roman{section}}
\renewcommand{\theequation}{A.\arabic{equation}}
\renewcommand{\thefigure}{A.\arabic{figure}}

\section{Upper bound efficiency of the protocol}
\label{sec:success}

Photons are distributed uniformly and randomly with no possibility of interference. In the case of two photons and eight parties, there are 64 possible configurations, as each photon can be sent independently to any of the eight parties. However, if the photons are sent to different parties, then once the first photon is assigned to a party, the second photon has 7 possible choices. Consequently, the total number of valid distributions is 56. Therefore, the probability that the two photons go to different parties is $56/64=0.875$ in the absence of experimental errors.

Another way to compute this probability is that there are only 8 cases where both photons go to the same party out of 64 possible configurations. Thus, the probability that both photons end up in the same party is $1/8=0.125$.

Generalizing the probability of photons going to different parties for an arbitrary $N$ and $K$, is $\frac{K!}{(K-N)!K^N}$
\section{Decoupling the Antimony Nucleus from the Emitted Photon}\label{decouplingSb}

To decouple antimony from the photon and obtain a pure photonic state, we apply a qudit Hadamard to the resultant state $\ket{\psi_{\text{final}}}$ and then measure the antimony.
Measuring the antimony is a common experimental process: it is done by sequentially applying an ESR pulse conditional on the nuclear states and flipping the spin state of the electron from $\ket{\downarrow}$ to $\ket{\uparrow}$. Then, readout the electron via spin to charge conversion. This fundamental process is performed on many other experiments. See details in\cite{Morello_2010, Yu2025, Stemp2024}. 

The process starts after the state in Eq: 2 in main text is created. The first step is to apply a qudit Hadamard to the resultant state after the time-bin multiplexing protocol. 
\begin{equation}
    \label{eq:quditH}
    \begin{aligned}
        H_8\ket{\psi_{\text{final}}} = (\ket{7/2} + \ket{5/2} +\ket{3/2} +\ket{1/2} +\ket{-1/2} +\ket{-3/2} +\ket{-5/2} +\ket{-7/2}  ) \otimes \ket{\downarrow} \otimes \ket{\omega_1}_{t_1}\\ 
        +(\ket{7/2} + \ket{5/2} +\ket{3/2} -\ket{1/2} -\ket{-1/2} -\ket{-3/2} -\ket{-5/2} +\ket{-7/2}  ) \otimes \ket{\downarrow} \otimes \ket{\omega_1}_{t_2}\\ 
        +(\ket{7/2} + \ket{5/2} -\ket{3/2} -\ket{1/2} +\ket{-1/2} +\ket{-3/2} -\ket{-5/2} -\ket{-7/2}  ) \otimes \ket{\downarrow} \otimes \ket{\omega_1}_{t_3}\\ 
        +(\ket{7/2} - \ket{5/2} -\ket{3/2} +\ket{1/2} -\ket{-1/2} +\ket{-3/2} +\ket{-5/2} -\ket{-7/2}  ) \otimes \ket{\downarrow} \otimes \ket{\omega_1}_{t_4}\\ 
         +(\ket{7/2} -\ket{5/2} +\ket{3/2} -\ket{1/2} +\ket{-1/2} -\ket{-3/2} +\ket{-5/2} -\ket{-7/2}  ) \otimes \ket{\downarrow} \otimes \ket{\omega_1}_{t_5}\\ 
         +(\ket{7/2} -\ket{5/2} +\ket{3/2} +\ket{1/2} -\ket{-1/2} +\ket{-3/2} -\ket{-5/2} -\ket{-7/2}  ) \otimes \ket{\downarrow} \otimes \ket{\omega_1}_{t_6}\\ 
         +(\ket{7/2} -\ket{5/2} -\ket{3/2} +\ket{1/2} +\ket{-1/2} -\ket{-3/2} -\ket{-5/2} +\ket{-7/2}  ) \otimes \ket{\downarrow} \otimes \ket{\omega_1}_{t_7}\\
         +(\ket{7/2} +\ket{5/2} -\ket{3/2} -\ket{1/2} -\ket{-1/2} -\ket{-3/2} +\ket{-5/2} +\ket{-7/2}  ) \otimes \ket{\downarrow} \otimes \ket{\omega_1}_{t_8}\\
    \end{aligned}
\end{equation} where $H_8$ represents the qudit Hadamard matrix for SU(8).
If we re-group the terms:
\begin{equation}
   \label{eq:regrouped}
    \begin{aligned}
        H_8\ket{\psi_{\text{final}}} = \ket{7/2}\otimes \ket{\downarrow} \otimes (\ket{w_1}_{t_1} +\ket{w_1}_{t_2} + \ket{w_1}_{t_3} + \ket{w_1}_{t_4} + \ket{w_1}_{t_5} + \ket{w_1}_{t_6} +\ket{w_1}_{t_7} + \ket{w_1}_{t_8})\\
        +\ket{5/2}\otimes \ket{\downarrow} \otimes (\ket{w_1}_{t_1} +\ket{w_1}_{t_2} + \ket{w_1}_{t_3} - \ket{w_1}_{t_4} - \ket{w_1}_{t_5} - \ket{w_1}_{t_6} - \ket{w_1}_{t_7} + \ket{w_1}_{t_8})\\
        + \ket{3/2}\otimes \ket{\downarrow} \otimes (\ket{w_1}_{t_1} +\ket{w_1}_{t_2} - \ket{w_1}_{t_3} - \ket{w_1}_{t_4} + \ket{w_1}_{t_5} + \ket{w_1}_{t_6} -\ket{w_1}_{t_7} - \ket{w_1}_{t_8})\\
        + \ket{1/2}\otimes \ket{\downarrow} \otimes (\ket{w_1}_{t_1} -\ket{w_1}_{t_2} - \ket{w_1}_{t_3} + \ket{w_1}_{t_4} - \ket{w_1}_{t_5} + \ket{w_1}_{t_6} +\ket{w_1}_{t_7} - \ket{w_1}_{t_8})\\
        + \ket{-1/2}\otimes \ket{\downarrow} \otimes (\ket{w_1}_{t_1} - \ket{w_1}_{t_2} + \ket{w_1}_{t_3} - \ket{w_1}_{t_4} + \ket{w_1}_{t_5} - \ket{w_1}_{t_6} +\ket{w_1}_{t_7} - \ket{w_1}_{t_8})\\
        + \ket{-3/2}\otimes \ket{\downarrow} \otimes (\ket{w_1}_{t_1} - \ket{w_1}_{t_2} + \ket{w_1}_{t_3} + \ket{w_1}_{t_4} - \ket{w_1}_{t_5} + \ket{w_1}_{t_6} - \ket{w_1}_{t_7} - \ket{w_1}_{t_8})\\
        + \ket{-5/2}\otimes \ket{\downarrow} \otimes (\ket{w_1}_{t_1} -\ket{w_1}_{t_2} - \ket{w_1}_{t_3} + \ket{w_1}_{t_4} + \ket{w_1}_{t_5} - \ket{w_1}_{t_6} -\ket{w_1}_{t_7} + \ket{w_1}_{t_8})\\
        + \ket{-7/2}\otimes \ket{\downarrow} \otimes (\ket{w_1}_{t_1} +\ket{w_1}_{t_2} - \ket{w_1}_{t_3} - \ket{w_1}_{t_4} - \ket{w_1}_{t_5} - \ket{w_1}_{t_6} +\ket{w_1}_{t_7} + \ket{w_1}_{t_8})
    \end{aligned}
\end{equation}
The states of antimony and the time-bins are now in a product state instead of an entangled state.
If we now measure the nuclear states, we obtain the $\ket{W_8}$ state, which is:
\begin{equation}
    \ket{W_8} = \ket{w_1}_{t_1} + \ket{w_1}_{t_2} + \ket{w_1}_{t_3} + \ket{w_1}_{t_4} + \ket{w_1}_{t_5} + \ket{w_1}_{t_6} +\ket{w_1}_{t_7} + \ket{w_1}_{t_8}
\end{equation} up to a correction.
Measuring the state of the antimony is a simple experimental process: this is done by sequentially applying an ESR pulse conditional on the nuclear states, and flipping the spin state of the electron from $\ket{\downarrow}$ to $\ket{\uparrow}$. Then, readout the electron via spin to charge conversion. This fundamental process is performed on many other experiments. See details in\cite{Morello_2010, Yu2025, Stemp2024}.
Note that, for the sake of simplicity, we did not write the actual qudit Hadamard matrix. Equations~\ref{eq:quditH} and \ref{eq:regrouped} are up to a renormalization constant. 
The qudit Hadamard matrix is as follows:
\begin{equation}\label{gen_H_qudit}
\scalemath{0.85}{
    \begin{pmatrix}
          1/4b-1/4b^3&   1/4b-1/4b^3&   1/4b-1/4b^3&  1/4b-1/4b^3&  1/4b-1/4b^3&  1/4b-1/4b^3&   1/4b-1/4b^3&  1/4b-1/4b^3 \\
    1/4b-1/4b^3&         1/4+1/4a&  1/4b+1/4b^3&     \
   -1/4+1/4a&  -1/4b+1/4b^3&        -1/4-1/4a&  -1/4b-1/4b^3&         1/4-1/4a \\
     1/4b-1/4b^3&   1/4b+1/4b^3&  -1/4b+1/4b^3& -1/4\
b-1/4b^3&   1/4b-1/4b^3&   1/4b+1/4b^3& -1/4b+1/4\
b^3& -1/4b-1/4b^3 \\
    1/4b-1/4b^3&        -1/4+1/4a&  -1/4b-1/4b^3&      \
    1/4+1/4a&  -1/4b+1/4b^3&          1/4-1/4a&   1/4b+1/4\
b^3&        -1/4-1/4a \\
    1/4b-1/4b^3&  -1/4b+1/4b^3&  1/4b-1/4b^3&  -1/4\
b+1/4b^3&   1/4b-1/4b^3&  -1/4b+1/4b^3&  1/4b-1/4\
b^3&  -1/4b+1/4b^3 \\
    1/4b-1/4b^3&        -1/4-1/4a&  1/4b+1/4b^3&     \
    1/4-1/4a& -1/4b+1/4b^3&          1/4+1/4a& -1/4b-1/4\
b^3&         -1/4+1/4a \\
     1/4b-1/4b^3& -1/4b-1/4b^3&  -1/4b+1/4b^3&  1/4\
b+1/4b^3&  1/4b-1/4b^3&  -1/4b-1/4b^3& -1/4b+1/4\
b^3&  1/4b+1/4b^3 \\
     1/4b-1/4b^3&          1/4-1/4a&  -1/4b-1/4b^3&     \
   -1/4-1/4a&  -1/4b+1/4b^3&        -1/4+1/4a&  1/4b+1/4\
b^3&        1/4+1/4a 

    \end{pmatrix}
    }
\end{equation}
where $a = E(4) = i$ and $b = E(8)= 0.707 + 0.707i$
In this work, GAP (Groups, Algorithms, and Programming)~\cite{GAP4} software is used to compute Clifford matrices for $SU(8)$ - including the Hadamard gate. 
\section{Background Information for Third Quantisation}\label{background_3rdq}
In this section, we provide the details of the third quantisation framework from~\cite{rudolph2021everywhere}. We start by briefly examine first and second quantization and how these are combined into a third quantization picture. 

In the second-quantization picture each system corresponds to a mode, and the number of photons within a mode (the occupation number) is the state of the system~\cite{rudolph2021everywhere}. In first-quantization the systems are the photons, and the internal level occupations of the photons (the states) are the modes. We follow Rudolph’s notation and let $| \cdot\succ$ denote the states described in second quantization. Having four photons in four modes  with exactly one photon per mode, in the second-quantization picture corresponds to $\mid 1_a1_b 1_c 1_d\succ$, where the subscripts represent the modes. In the first-quantized picture, the state transforms as:
\begin{equation}\label{firstq}
   \begin{aligned}
       \mid 1_a 1_b 1_c 1_d\succ \Leftrightarrow \ket{\sigma_{1111}} = &\ket{a} \ket{b} \ket{c} \ket{d} +
\ket{a} \ket{b} \ket{d} \ket{c} +\\
&\ket{a} \ket{c} \ket{b} \ket{d} +
\ket{a} \ket{c} \ket{d} \ket{b} +\\
&\cdots+\\
&\ket{d} \ket{c} \ket{a} \ket{b} +
\ket{d} \ket{c} \ket{b} \ket{a}
   \end{aligned}
\end{equation} i.e. when describing the system in first quantization, we need to create an equal superposition over all permutations of the modes to account for the fact that photons are bosons and therefore the state is invariant when photons are interchanged. Describing state evolution is easier in the first-quantization picture. For example, sending four photons through an interferometer that performs a mode-transformation $U$ results in $(U \otimes U \otimes U \otimes U)\ket{\sigma_{1111}}$. However, measuring one mode in first-quantization is non trivial because it involves the wavefunction of all systems (photons), while in second-quantization, the measurement affects only the relevant system (mode).
If we could independently measure systems (photons) within first quantization, the statistical outcomes would remain unchanged, provided all measurements were identical. 
However, we cannot independently access the systems (photons) within first quantization. And this is precisely where third quantization becomes relevant. By embedding first-quantization within second-quantization — a method known as third quantization - we can access photons independently. In this approach, we uniformly and randomly distribute $N$ photons over $K$ parties, with $K \gg N$. Because $K$ is much larger than $N$, the probability of multiple photons going to the same party is reduced. If a single photon is spread over $K \gg N$ modes, the resultant state $\ket{W_K}$ can be written as follows in the second quantized picture:
\begin{equation}
    \begin{aligned}
        \ket{W_K} &= \frac{1}{\sqrt{K}} \sum_{J=1}^K \mid 1_j\succ\\
                  &=\frac{1}{\sqrt{K}}(\mid 100\cdots0\succ +\mid 010\cdots\succ + \mid 000\cdots1\succ)
    \end{aligned}
\end{equation} where each mode is sent to one of the parties. If we repeat this process with $N$ copies of $\ket{W_K}$, then each party will hold $N$ modes.  We focus on the case of $N=4$ photons distributed over $K$ parties. 
\begin{equation} \label{eq:3q}
   \begin{aligned} \ket{W_K}^{\otimes 4} \approx \ket{\Sigma_{ABCD}} + \ket{\Sigma_{ABCE}} + \cdots + \ket{\Sigma_{VXYZ}} 
   \end{aligned} 
\end{equation} where the capital letters represent the parties holding the photons, and there are $\frac{K!}{(K-4)!}$ many permutations. We use $\approx$ because we disregard the parts of the wavefunction of $\ket{W^4}$ where an individual party has more than one photon. If $K = N^3$, then the error in this approximation scales as $O(1/N)$ as $ N \rightarrow \infty$, meaning that we can bound the error arbitrarily small by using more photons~\cite{rudolph2021everywhere}. 
If the four-photon state is held by $V, X, Y, Z$, then $\ket{\Sigma_{VXYZ}}$ is as follows: 
\begin{equation}\label{thirdq}
   \small
   \begin{aligned}
     \ket{\Sigma_{VXYZ}} = \frac{1}{\sqrt{4!}} & (\mid1000\succ_V \mid0100\succ_X \mid0010\succ_Y \mid0001\succ_Z + \\
     & \mid1000\succ_V \mid0100\succ_X \mid0001\succ_Y \mid0010\succ_Z + \\
     &\cdots + \\
     & \mid0001\succ_V \mid0010\succ_X \mid0100\succ_Y \mid1000\succ_Z )
   \end{aligned}
\end{equation}
There is a similarity between state $\ket{\sigma_{1111}}$ in Eq.~\ref{firstq} and state $\ket{\Sigma_{VXYZ}}$ (though they are not exactly equal to each other, as the number of modes is different).
The state $\ket{\Sigma_{VXYZ}}$ is written in the second-quantization picture but it encodes a maximally-symmetric first-quantized state of four photons where the parties $V, X, Y, Z$ hold a single photon  in four  modes within their respective labs. This is what we call a third quantized state~\cite{rudolph2021everywhere}. The $\ket{\Sigma_{VXYZ}}$ is a four photon state but in sixteen modes and not in four modes. Unlike $ \mid 1_a 1_b 1_c 1_d\succ$, the state $\ket{\Sigma_{VXYZ}}$ is an entangled state. We can consider a specific type of measurement for the state $\ket{\Sigma_{VXYZ}}$ where each party performs photodetection on the same mode—for example, photodetection on the first mode they hold without interacting with the other three modes. 
Depending on the unitary applied before the measurement, a certain number $n$ of the parties will observe the state $\ket{a}$. This corresponds to the detection of $n$  photons in mode $a$, which is a projection onto $\mid{n_a}\succ$. The implication of this is that four parties possessing the state $ \ket{\Sigma_{VXYZ}} $ are capable of simulating any arbitrary protocol on four single photons. Because, the evolution of a qudit encoded via a single photon in $d$ modes can be fully simulated using only linear optics, whatever evolution $ U$  describes the interferometer through which the photons pass, parties $V$, $X$, $Y$ and $Z$ can independently implement it on their respective modes. This results in the $ U \otimes  U \otimes U \otimes U$ evolution that replicates the first-quantized scenario. Furthermore, one can define a specific type of
measurements in which each party performs independent measurements on the subset of modes they hold, followed by feedforward (i.e., outcome-dependent interferometers
on the remaining unmeasured modes prior to their measurement~\cite{rudolph2021everywhere}. For example, if each party holds four photons in total, they may choose to measure only the first photon
while leaving the remaining three untouched. All photons undergo identical measurement procedures at each step. In most systems, this process yields a rank $> 1$ outcome, leaving the measured photons still entangled with the unmeasured ones.) As it was shown in~\cite{rudolph2021everywhere, Knill2001}, this is universal for quantum computation The only necessity for this protocol is classical communication and local unitary operations. 
The third quantized state $\ket{\Sigma_{VXYZ}}$ is highly entangled and created without the use of any two-qubit gates or interactions between individual photons~\cite{Gross_PRL_2009, Raussendorf_2001}.The details of the third quantisation framework, as well as how universal quantum computing can be achieved using third quantisation, can be found in~\cite{rudolph2021everywhere}.

\section{Hamiltonian for cavity coupled with Antimony}\label{total_H}
For time-bin multiplexing, we utilize a single microwave cavity fabricated on our silicon chip, which is coupled to an antimony donor. This cavity is designed to operate at the EDSR frequency, specifically between the states  $\ket{\downarrow}\ket{7/2} \leftrightarrow \ket{\uparrow}\ket{5/2}$. The total Hamiltonian of the system is given by:
\begin{equation}
    H_{\text{total}} = H_{Sb} + H_{\text{interaction}} + H_{\text{field}}
\end{equation}
where $H_{Sb}$ is the Hamiltonian of the donor which is defined in Eq~1 in main text. $ H_{\text{interaction}} + H_{\text{field}}$ part represents the emitting photon through cavity. The total Hamiltonian can be expressed as follows:
\begin{equation}\label{drift_Ham}
    \begin{aligned}
        H_{\text{total}} =  \underbrace{ B_0(-\gamma_n \hat{I_z} + \gamma_e \hat{S_z}) + A(\Vec{S}\cdot\Vec{I}) + \Sigma_{\alpha, \beta \in \{x,y,z\}} Q_{\alpha \beta}\hat{I}_\alpha \hat{I}_\beta}_{H_{Sb}} \\
        + \left (\underbrace{ \hbar w_{\ket{7/2}\leftrightarrow\ket{5/2}} a^{\dag} a}_{H_{\text{field}}}       
        +   \underbrace{g (\ket{7/2}\ket{\downarrow}\bra{5/2}\bra{\uparrow} a^{\dag}) + g(\ket{5/2}\ket{\uparrow}\bra{7/2}\bra{\downarrow} a)}_{H_{\text{interaction}}} \right)\\       
    \end{aligned}
\end{equation}
 where the operators $a^{\dag}$ and $a$ represent creation and annihilation operators, respectively. $g$ represents the coupling strength between the spin and the cavity. When purely magnetic coupling is employed via the ESR transitions, $g$ is in the range of $10$Hz - $100$kHz~\cite{Wyatt_1}. However, when utilizing an electric dipole for spin-cavity coupling, significantly higher coupling strengths -- in the range of a few MHz\cite{tosi_2017} -- and thus faster photon emission times are possible. This is why we propose to operate the cavity at the antimony donor EDSR transitions.
\section{Frequency Multiplexing Protocol}\label{freq-mult}
\begin{figure*}[htb!]
    \includegraphics[width = .9\linewidth]{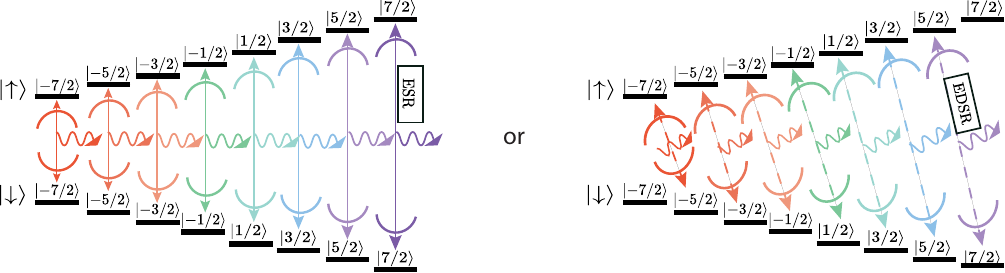}
    \caption[Frequency multiplexing protocol]{\label{fig:freq}\textbf{Frequency multiplexing protocol} The protocol uses as many microwave cavities as there are frequencies in the system (eight microwave cavities for ESR frequencies and seven microwave cavities for EDSR frequencies).
    }
\end{figure*}
We start by considering the use of all the ESR frequencies, utilizing eight cavities corresponding to these frequencies. The protocol unfolds as follows: 

i) Generate an equal superposition of all states in the down subspace of the electron, denoted as $\ket{\downarrow}$. This can be achieved by applying a qudit Hadamard operation on the initial state we choose. For example, if the initial state is chosen as $\ket{7/2}\ket{\downarrow}\ket{vac}$, the state after the qudit Hadamard operation is applied becomes: \[
    \begin{split}
        \ket{\psi_1}=\frac{1}{\sqrt{8}}\big(& \ket{7/2} + \ket{5 /2} + \ket{3/2} \\&+ \ket{1/2} + \ket{-1/2} + \ket{-3/2} \\&+ \ket{-5/2} + \ket{-7/2} \big) \ket{\downarrow}\ket{vac},
    \end{split}
\] where the $\ket{vac}$ represents the vacuum and electron is in the spin-down state. The first step is the same as in the time-bin multiplexing protocol. 

ii) Apply a broad range frequency spectrum which includes all eight frequencies to flip the spin of the electron from $\ket{\downarrow}$ to $\ket{\uparrow}$ and to bring the equally superposed state to the electron $\ket{\uparrow}$ subspace. The state then becomes: \[
    \begin{split}
        \ket{\psi_2}=\frac{1}{\sqrt{8}}\big(& \ket{7/2} + \ket{5 /2} + \ket{3/2} \\&+ \ket{1/2} + \ket{-1/2} + \ket{-3/2} \\&+ \ket{-5/2} + \ket{-7/2} \big) \ket{\uparrow}\ket{vac},
    \end{split}
\]

iii) Allow the system to relax and emit a photon through one of the eight cavities. Here, each cavity can emit a single photon at a time. The antimony donor has one extra electron that is weakly coupled to it, allowing the emission of only one photon. The emitted photon will be in a superposition of all eight frequencies. The resultant state will be a $W$ state, which has 8 terms, each corresponding to a frequency of the antimony donor. 
\begin{equation}
    \begin{aligned}
        \ket{\psi_f} = \frac{1}{\sqrt{8}}(\ket{7/2}\ket{\downarrow}\ket{\omega_1}+ \ket{5/2}\ket{\downarrow}\ket{\omega_2} + \\ \cdots + \ket{-7/2}\ket{\downarrow}\ket{\omega_8})
    \end{aligned}
\end{equation}

We can represent the frequencies in binary form such as $\ket{\omega_1} = \ket{10000000}$ and $\ket{\omega_2} = \ket{01000000}$ and so on. The degree of freedom of the protocol is the frequencies. The application of an additional Hadamard operation on the state $\ket{\psi_f}$, followed by measuring the antimony, will result in a state $\ket{W_8}$ (see details in \ref{decouplingSb}). The total Hamiltonian of the system is:
\begin{equation}\label{drift_H}
    \begin{aligned}
        H_{\text{total}} =  \underbrace{ B_0(-\gamma_n \hat{I_z} + \gamma_e \hat{S_z}) + A(\Vec{S}\cdot\Vec{I}) + \Sigma_{\alpha, \beta \in \{x,y,z\}} Q_{\alpha \beta}\hat{I}_\alpha \hat{I}_\beta}_{H_{Sb}} \otimes I^{\otimes 8}\\
        + I^{\otimes 16} \otimes \left (\underbrace{\sum_{i=1}^{i=8} w_i a_{i}^{\dag} a_i}_{H_{\text{field}}}       
        +   \underbrace{\sum_{i=1}^{i=8} g_i (\ket{i}\ket{\downarrow}\bra{i}\bra{\uparrow} a_i^{\dag}) + g_i (\ket{i}\ket{\uparrow}\bra{i}\bra{\downarrow} a_i)}_{H_{\text{interaction}}} \right)\\       
    \end{aligned}
\end{equation} where the operators $a_i^{\dag}$ and $a_i$ represent creation and annihilation operators, respectively. Specifically, $a_1 = a \otimes I \otimes I \cdots = a \otimes I^{\otimes 7}$ and $a_2 = I \otimes a \otimes I \cdots = I \otimes a \otimes I^{\otimes 6}$, and so forth, where $a=\begin{pmatrix} 0  0 \\ 1 0 \end{pmatrix}$. The $g_i$ terms represent the coupling strength between the spin and the cavity. As we have eight distinctly different frequencies, the $g_i$ values are all unique. 

There are a few concerns with this protocol. 
Fabricating eight cavities onto one antimony device is not feasible with current technology. By using EDSR frequencies instead of ESR, the number of cavities can be reduced to seven while the resulting state remains unchanged, ensuring the formation of $W$ states at the end of the protocol. In this case, a photon is spread over seven modes instead of eight, hence the resultant state becomes $\ket{W_7}$ instead of $\ket{W_8}$. However, fabricating seven cavities on one chip and coupling them to a single antimony donor spin is still highly impractical. In addition, the coupling strength of each of the seven cavities to the antimony donor would need to be identical in order to create an entangled $W$ state. To address this critical challenge, the utilization of tunable cavities emerges as a viable solution. In both ESR and EDSR applications, the frequencies of the device deviate by approximately $100$MHz between consecutive transitions. Consequently, a tunable cavity with an $800$ MHz frequency range (or $700$ MHz if EDSR is used) would be required. This approach could streamline the system by employing a single cavity, thereby ensuring a consistent coupling strength. 
However, making a cavity that can be tuned within a 700 MHz gap is also challenging. 
It is conceivable that future technological advancements may enable the fabrication of devices incorporating seven cavities, obviating the need for tunable cavities, or potentially enabling the use of a tunable cavity that can be adjusted up to 700 MHz, along with a feasible readout mechanism. Until such advancements materialize, we propose a different version of this approach herein, namely, time-bin multiplexing. 
\newpage
\section{Steps of Time-bin Multiplexing Protocol}\label{time-binsteps}
\begin{figure*}[htb!]
    \includegraphics[width = .9\linewidth]{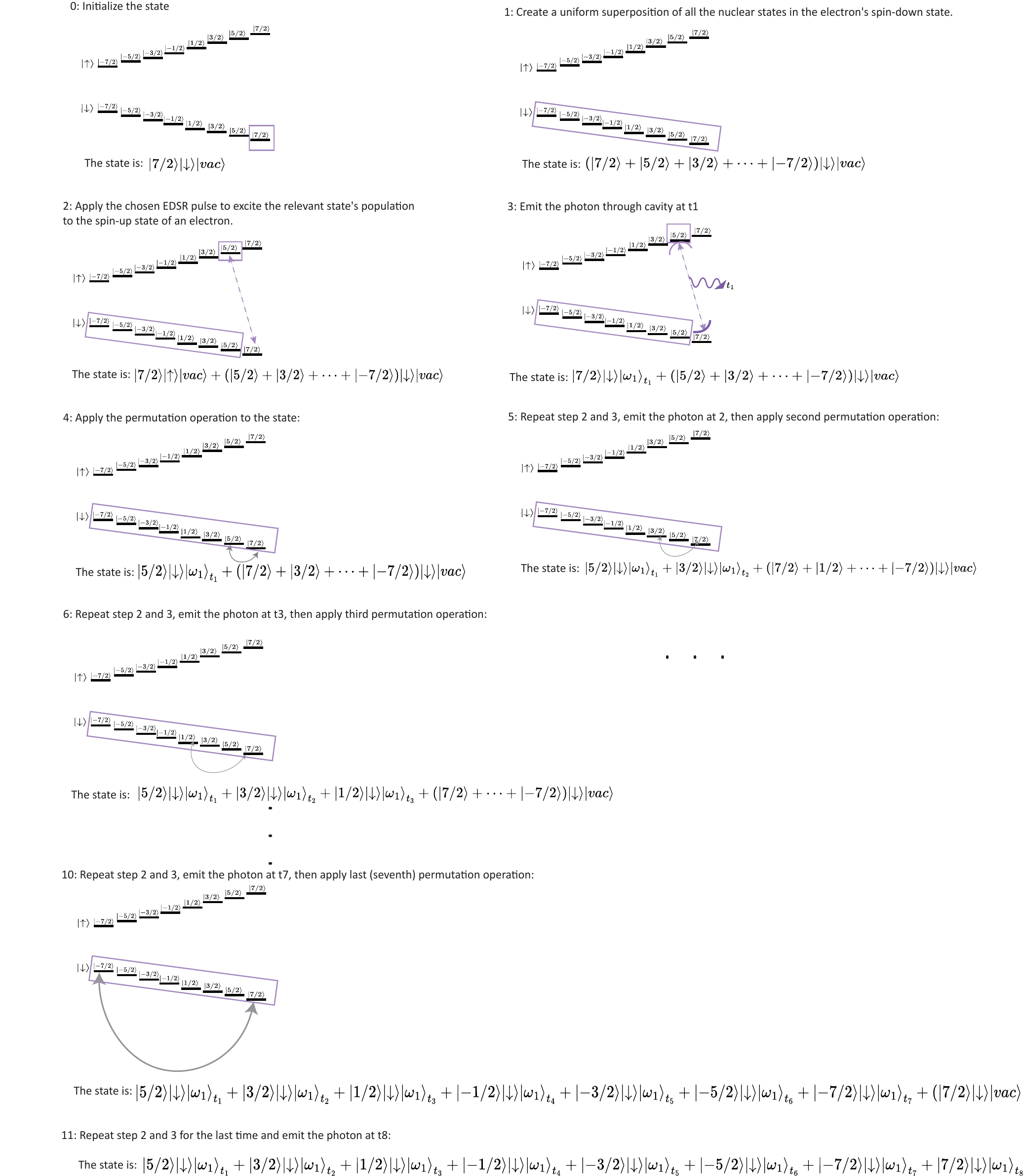}
\end{figure*}
where normalization factor is neglected.



\newpage
%
\end{document}